\documentclass[a4paper,12pt]{article}
\usepackage{textcomp}
\usepackage{graphicx}
\usepackage{amsfonts}
\usepackage{amsmath}
\usepackage{amsthm}
\usepackage{amssymb}
\usepackage{url} 
\usepackage[all]{xy}
\usepackage{fancyhdr}
\usepackage[none]{hyphenat}

\usepackage{array} 
\usepackage{hyperref}
\hypersetup{
    colorlinks=true,
    linkcolor=blue,
    filecolor=magenta,      
    urlcolor=cyan,
    pdftitle={20191128},
    pdfpagemode=FullScreen,
    }

\usepackage{float}
\floatstyle{plaintop}

\usepackage{afterpage}

\usepackage[active]{srcltx}
\usepackage{times}
\usepackage{caption}
\usepackage{comment}
\usepackage[margin=1in]{geometry}

\usepackage{graphicx, wrapfig, subcaption, setspace, booktabs}
\usepackage[T1]{fontenc}
\begin{document}

\title{Multivariate data analysis using recurrence measures}  

\author{
\small \textbf{Shivam Kumar}$^{1}$, \textbf{R. Misra}$^{2}$, and \textbf{G. Ambika}$^{3}$ \\
\small $^{1}$Indian Institute of Science Education and Research (IISER) Tirupati, Tirupati-517 619, India \\
\normalsize $^{2}$Inter-University Centre for Astronomy and Astrophysics (IUCAA), Pune-411 007, India \\
\small $^{3}$Indian Institute of Science Education and Research (IISERTVM), Thiruvananthapuram-695 551, India
}

\date{}  

\maketitle
.
\begin{abstract}
The emergent dynamics of complex systems often arise from the  internal dynamical interactions among different elements and hence is to be modeled using multiple variables that represent the different dynamical processes. When such systems are to be studied using observational or measured data,  we may benefit from using data from all variables or observations of the system rather than using that from a single variable. In this study, we try to bring out the relative effectiveness of the analysis of data from multiple variables in revealing the underlying dynamical features.  For this, we derive the recurrence measures from the multivariate data of standard systems in periodic, chaotic and hyper chaotic states and compare them with that from noisy data.  We  identify Entropy computed from Recurrence Plot and Characteristic Path Length from recurrence network as the most effective measures that can identify the nature of the dynamical state of the system, and differentiate it from stochastic or noisy behaviour. We find that for different variables, the recurrence measures to be mostly similar for data from periodic states, while they differ for chaotic and hyperchaotic states, indicating that multi-variate analysis is useful for real world systems in the latter states.
\end{abstract}


\section*{Introduction}

  Studies to understand the dynamics of complex systems from their time-series data form an active area of recent research. Due to its wide range of applications in climate, astrophysics, finance, bio-medicine, neurology, etc., and the availability of large amount of data from such systems, new and potent data processing techniques are being developed. They help to categorize the body of knowledge available in datasets appropriately for further understanding and modeling.  The basic technique in this context follows a dynamical system's approach, where the variations in the data are attributed to the manifestations of the underlying dynamics. Using the well established technique of delay embedding, the dynamics is reconstructed and the patterns in the embedded phase space are extracted through measures that can capture the inherent structures hidden in the data. These measures can then be connected to details about the dynamics of the system producing the data and transitions that can happen in its dynamics over time \cite{10.1063/1.4917289, Ambika2020}. In this context, the well established method is recurrence quantification analysis(RQA) where the pattern of recurrences in the reconstructed trajectory is captured through recurrence plots and recurrence networks\cite{Marwan2015,ZOU20191,Marwan2023}. The measures computed from them, are most effective in extracting the dynamical features from the time series and are especially effective with short and non-stationary data\cite{MARWAN2007237}. Such feature extractions can further be of use with machine learning algorithms to manage large numbers of data sets efficiently \cite{10.1063/5.0196382}.
 
 \sloppy
 While extensive research on univariate time-series data from synthetic and real-world sources has flourished, the comparative analysis of data from different observables available from the same system or multivariate data generated by different processes in the same source still needs more attention. We come across such multivariate data in real complex systems such as EEG and ECG data from multiple channels, meteorological data of temperature, relative humidity, and pressure from one location, astrophysical data at different energy bands or wavelengths from one source, experimental or survey data that arise in turbulence, combustion, fluid flow, business, healthcare, and social sciences, economic data from stock markets and currencies etc. The analysis of multivariate data can be directly useful to understand the relationship between customer behavior and marketing campaigns in industry \cite{cleff_2019_applied}, the relationship between patient symptoms and diseases in healthcare, the influence of social factors in political outcomes in social sciences \cite{liao_2010_analysis}, the relative impact of different species in ecology \cite{babak_2024_stochastic}, etc.
 
\sloppy
 
 Some of the recent studies that utilize multivariate time series data report use of moving dynamic principal component analysis using MDPCA, a method that reduces the dimension of data by transforming it into uncorrelated components \cite{alshammri_2021_moving}, study in terms of ensembles of correlation matrices\cite{vyas_2018_multivariate} and analysis of the statistical properties of recurrence plots from time-frequency signal representations of biological signals \cite{10.1063/5.0026954}. The complexity of the human cardiac system is explored using multiplex recurrence networks, or Multivariate Joint Order Recurrence Networks (MJORN), mainly for diagnostics\cite{ssrn, 10.1063/5.0167477,PhysRevE.97.012312}. The multivariate data analysis is also used for detecting multiple change-points\cite{Min2014ChangepointDI} and for efficient segmentation using community detection  algorithm\cite{10.1063/5.0152881}. 
 Recently, Multidimensional Recurrence Quantification Analysis(MdRQA) is introduced as a tool to analyze multidimensional time series data that can be used to capture the dynamics of high-dimensional signals\cite{PMID:27920748} and later  its extension as Multidimensional Cross-Recurrence Quantification Analysis (MdCRQA)\cite{wallot2019multidimensional}. Also in this context, a new concept called nonlinear observability is reported that can determine if a dynamical system can be reconstructed from one signal or a combination of signals\cite{10.1063/1.5049903}. It is worth mentioning that  the relationship between the synchronicity of time series and the corresponding topological properties are captured through the cross-recurrence plots and joint recurrence plots that provide a general framework in the analysis of complex systems.\cite{Song_2024,ZIAEIHALIMEJANI2021107549}. The issue of choosing the  correct variables for reconstruction as well as the influence of symmetries in this choice are discussed in later studies\cite{10.1063/1.1487570}. In all these studies, multiple embeddings possible from combinations of variables or embeddings using all or combination data form the main focus. However embedding from data of each variable and its relative effectiveness and comparison across measures from different variables are not much explored.   

\sloppy

According to the Taken's embedding theorem, a single  time series or data from one variable should be enough for reconstructing the dynamics. However, in many cases, multiple data sets are available and then it is not clear which one can be most effective for modeling and prediction\cite{10.1371/journal.pone.0018295}. Also we may benefit from using all the available datasets of the system for better understanding of the underlying dynamics or validation of the results from one observable. Thus the analysis of multivariate data primarily helps to understand the relationships between variables, bringing out the dependence between their time series, and assessing the extend to which a given time series is similar or dissimilar to the others. In real complex systems, the dynamical processes or mechanisms that produce their time-series data are difficult to capture solely based on the values of a single observable or variable. Therefore, it is highly relevant to investigate the nonlinear measures obtained from the time series data of all the observables to arrive at the complex relationships and information flow among different components or variables from the same source.

In this study, we present a comprehensive analysis of the multivariate data generated from standard systems viz.  Rössler, Lorenz, and Chen systems and investigate the information content distribution across the different variables when their dynamical states are periodic, chaotic, or hyperchaotic. Most of the reported research on data from these systems uses the time series from one variable only. We consider the time series data of all the variables of each system and analyze the recurrence-based measures using the recurrence plot and recurrence networks generated from the reconstructed dynamics of each data. This could lead to a better understanding about which of the variables of the system can capture its dynamics most effectively and therefore will be most useful in predictions. This study holds particular relevance for domains where data acquisition spans multiple wavelengths, different energy bands, or channels.

\section*{Methodology}

We start by generating synthetic data from three standard systems, Rössler, Lorenz, and Chen systems from their dynamical equations numerically for chosen parameter values such that they correspond to different dynamical states like periodic, chaotic, and hyperchaotic. 
As part of pre-processing, we do a uniform deviate\cite{PhysRevE.93.012202} of each data set that ensures a more balanced representation that minimizes edge effects by standardizing the data range. The first step in the recurrence analysis is the reconstruction of dynamics from data for which we apply the Takens' delay embedding procedure. 
 For a given time series \(s(t)\), the reconstructed vector space \(\vec{x}(t)\) is represented by the equation below:
\begin{equation}
    \vec{x}(t) = (s(t), s(t+\tau), s(t+2\tau), ..., s(t+(m-1)\tau))
    \label{eq:recon}
\end{equation},
where \(\tau\) is the time delay, \(m\) is the embedding dimension, and  \(t\) denotes the time index. Following the standard procedure, we set the delay time (\(\tau\)) for each data set,  as \(1/e\) of the auto-correlation\cite{10.1063/5.0137223}. Since we are  using data from standard systems in the present study, we take the embedding dimension \((m)\) to be the actual dimension in each case.

The dynamics underlying data are often recurrent being confined to a region of the reconstructed state space. The recurrence pattern in the reconstructed trajectory is captured through the recurrence matrix defined as,
 
\begin{equation}
    R(i, j)=\Theta(\varepsilon-\|\vec{x}(i)-\vec{x}(j)\|)
    \label{eq:rec_mat}
\end{equation}

where \((i,j)\) are time indices. \(\epsilon\) denotes the chosen distance threshold , and \(\Theta\) the Heaviside function. 
Here again, the selection of \(\epsilon\), the distance threshold, is important for which we follow the method in \cite{10.1063/1.5024914} and take \(\epsilon\) to be the \(8^{th}\) percentile of the distribution of the pairwise distances in the reconstructed m-dimensional state space. The pattern of recurrences is then visualized as a 2-D Recurrence Plot (RP).

From the recurrence matrix, the adjacency matrix of the corresponding recurrence network(RN) is constructed as 

\begin{equation}
    A(i, j)(\varepsilon) = R(i, j)(\varepsilon)-\delta(i, j)
    \label{eq:adj_mat}
\end{equation}
By construction, the RN considers each point on the reconstructed trajectory as the node and links are made between pairs of points that lie within the threshold \(\epsilon\).  The Kronecker delta term, \(\delta(i,j)\), eliminates the self-loops in the network.

The recurrence quantification analysis (RQA) involves the computation of measures from the RP, that captures the distribution of dots, and diagonal lines as well as vertical and horizontal lines. The measures commonly used are Determinism (DET), Laminarity (LAM), Recurrence rate (RR) and Entropy (ENT).\cite{10.1063/1.4934554}  Similarly, the topology of RN that reflects the recurrence pattern is captured using measures of Link Density (LD), Global Clustering Coefficient (GCC), Characteristic Path Length (CPL), Network Entropy (NET ENT), Transitivity(TRA) and Assortativity (ASSOR)\cite{yang_2019_a}.
We note that most data sets from real-world systems are not large in size for various reasons like gathering large sets can be expensive and time-consuming, or data is often noisy, incomplete, or come with gaps.  Even when analysing synthetic data, large data sets can be computationally expensive and hence desirable to work with small data sets. So we check the robustness of all these measures for small data sets and find that the measures, DET, ENT, CPL, and GCC are robust and give consistent results even with small datasets of even 500 points.  Hence we compute mainly these four measures as detailed below and compare the results to see how they differ across multivariate data sets. 
\begin{itemize}

\item \textbf{Determinism (DET)}
quantifies occurrence of the diagonal line segments in RP and is defined as: 
\begin{equation}
    D E T=\frac{\sum_{l=l \min }^N l P(l)}{\sum_{l=1} l P(l)}
\end{equation}

where \(l\) stands for the lengths of diagonal lines in the RP, and \(P(l)\) is the number of lines of length equal to \(l\). We take \(l_{min}\) as the first maximum of \(l*P(l)\) vs. \(l\) \cite{10.1063/5.0165282}.
DET can be interpreted as the measure of predictability of the dynamics with high DET indicating deterministic and nonlinear nature.

\item \textbf{Entropy (ENT)}
refers to the Shannon entropy computed from the distribution of the diagonal line segments of lengths \(l\) in RP.
\begin{equation}
    E N T =-\sum_{l=l_{\min }}^N p(l) \ln p(l)
\end{equation}
ENT reflects the complexity of the deterministic structure in the system with uncorrelated noise giving very low values of ENT\cite{10.1371/journal.pone.0185968}.

\item \textbf{Characteristic Path Length (CPL)}
is defined for the recurrence network as the average over the shortest distances for every pair of nodes in the network:
\begin{equation}
    C P L=\frac{1}{N(N-1)} \sum_{i \neq j}^N l_{i j}
\end{equation}
where \(l_{ij}\) is the shortest distance (minimum number of links) between nodes i and j. CPL indicates the complexity of the system's behavior, with higher CPL indicating more complex and irregular behavior\cite{M2022128240}.

\item \textbf{Global Clustering Coefficient (GCC)}
 is computed as the average of the local clustering coefficients.The local clustering coefficient, \(C_{i}\),  for a node \(i\) in the network is defined as the ratio of the number of triads it is part of to the number of such possible triads, including node \(i\) \cite{Donner2015}. 
\begin{equation}
GC C=\frac{\sum_{i=1}^N C_i}{N} ; \text{ where }
C_i=\frac{\sum_{j, q} A_{i j} A_{j q} A_{q i}}{k_i\left(k_i-1\right)} 
\end{equation}

Here \(k_{i}\) is the degree of node {i}. GCC is a good indicator for the presence of regular dynamics underlying the data. \cite{MARWAN20094246}
\end{itemize}

In the following sections, we will present the computed values of these measures for data sets of different types of dynamical states, like  periodic, chaotic, hyperchaotic and see how they can distinguish deterministic and noisy behaviour.

\subsection*{Recurrence measures from periodic, chaotic and hyperchaotic data sets.}

The equations representing the dynamics of the chosen three standard nonlinear systems and the parameter values used in each case to get periodic, chaotic and hyperchaotic data sets are given in Table \ref{table:systems}.  

\begin{figure}[H]
    \centering
    \includegraphics[width=1\linewidth]{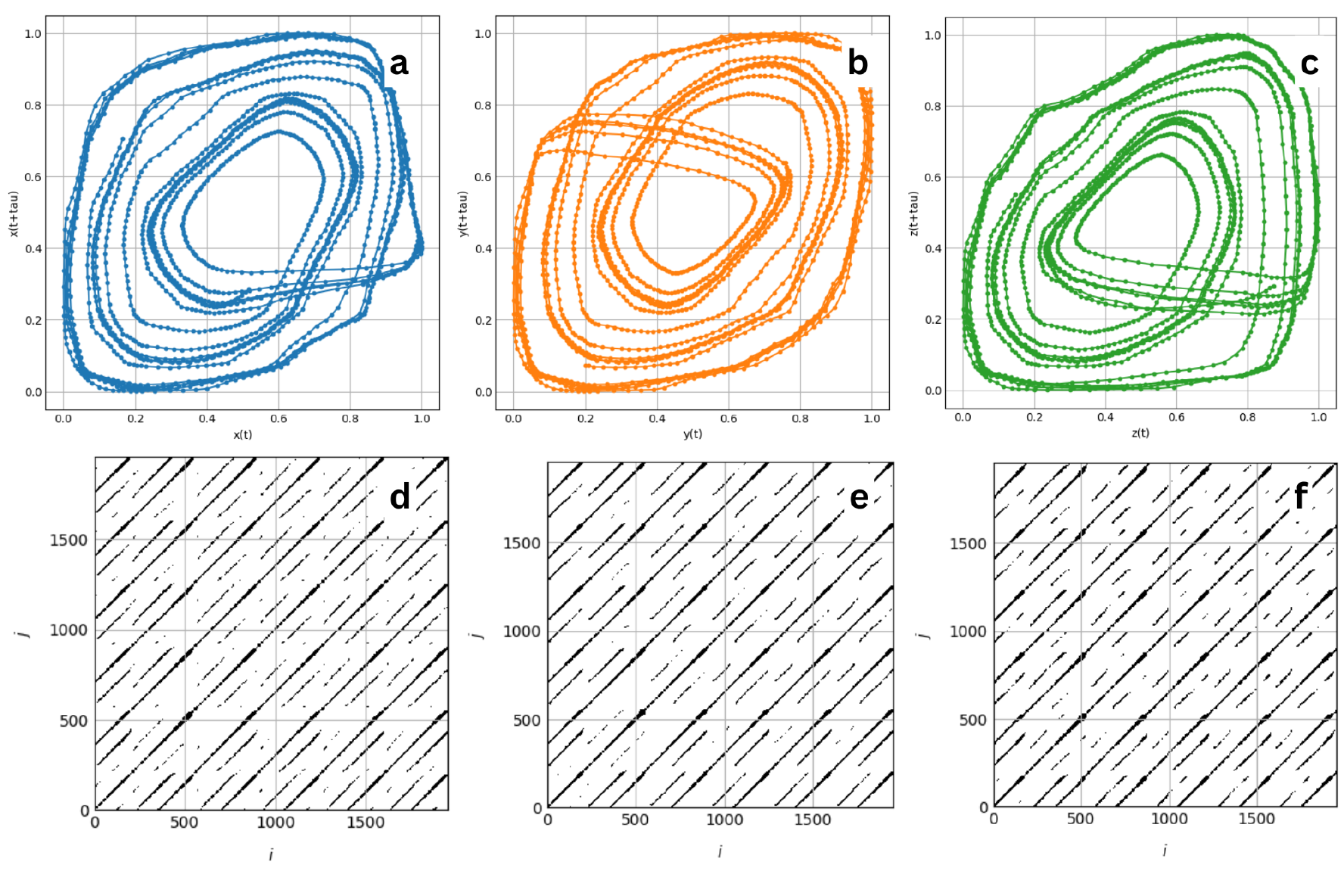}
    \caption{Reconstructed phase space structure and the corresponding Recurrence Plots from data of the Rössler system in chaotic region for (a,d) x-data, (b,e) y-data, and (c,f) z-data.}
    \label{fig:rossler}
\end{figure}

\begin{figure}[H]
    \centering
    \includegraphics[width=1\linewidth]{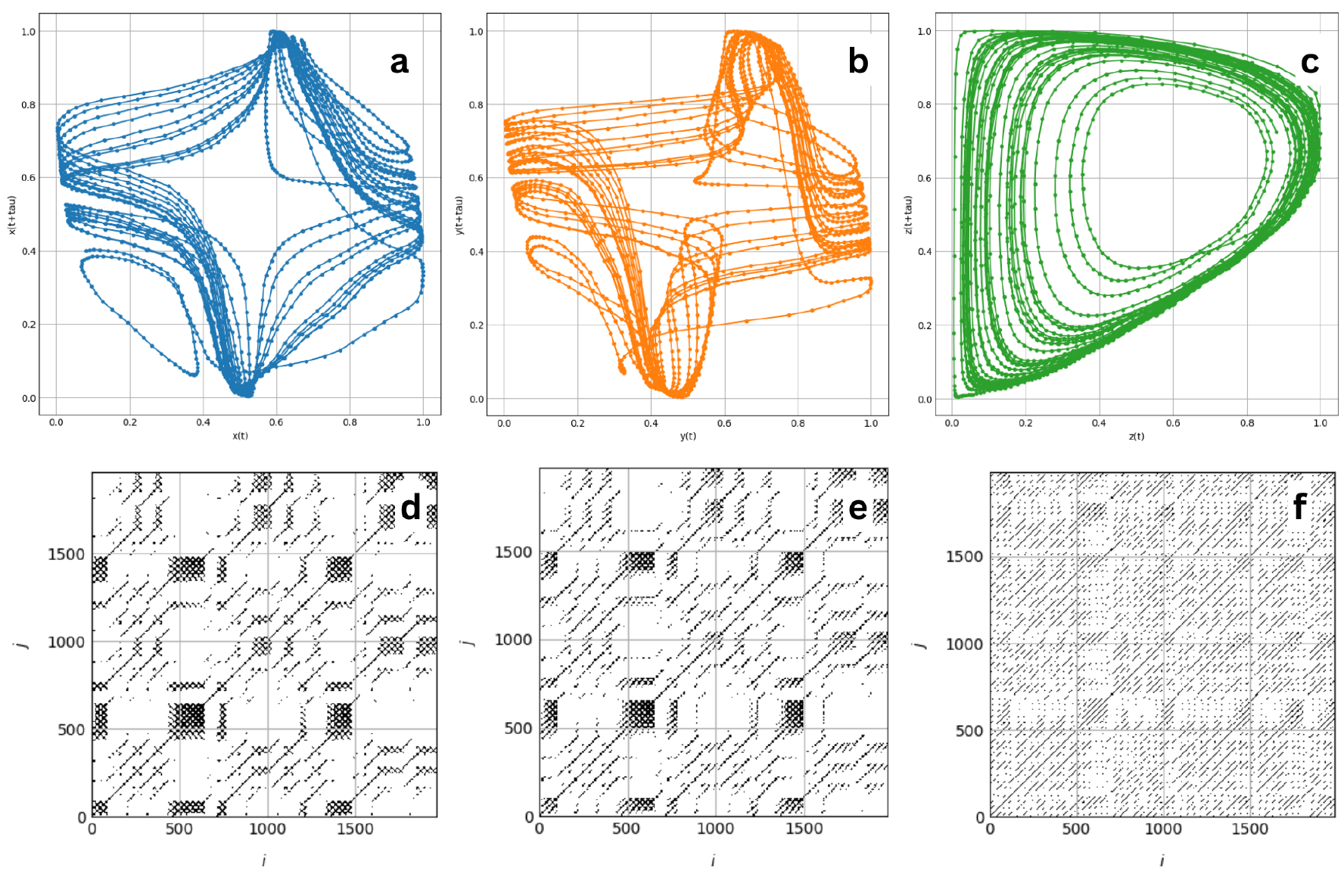}
    \caption{Reconstructed phase space structure and the corresponding Recurrence Plots from data of the Lorenz system in chaotic region for (a,d) x-data, (b,e) y-data, and (c,f) z-data.}
    \label{fig:lorenz}
\end{figure}

\begin{figure}[H]
    \centering
    \includegraphics[width=1\linewidth]{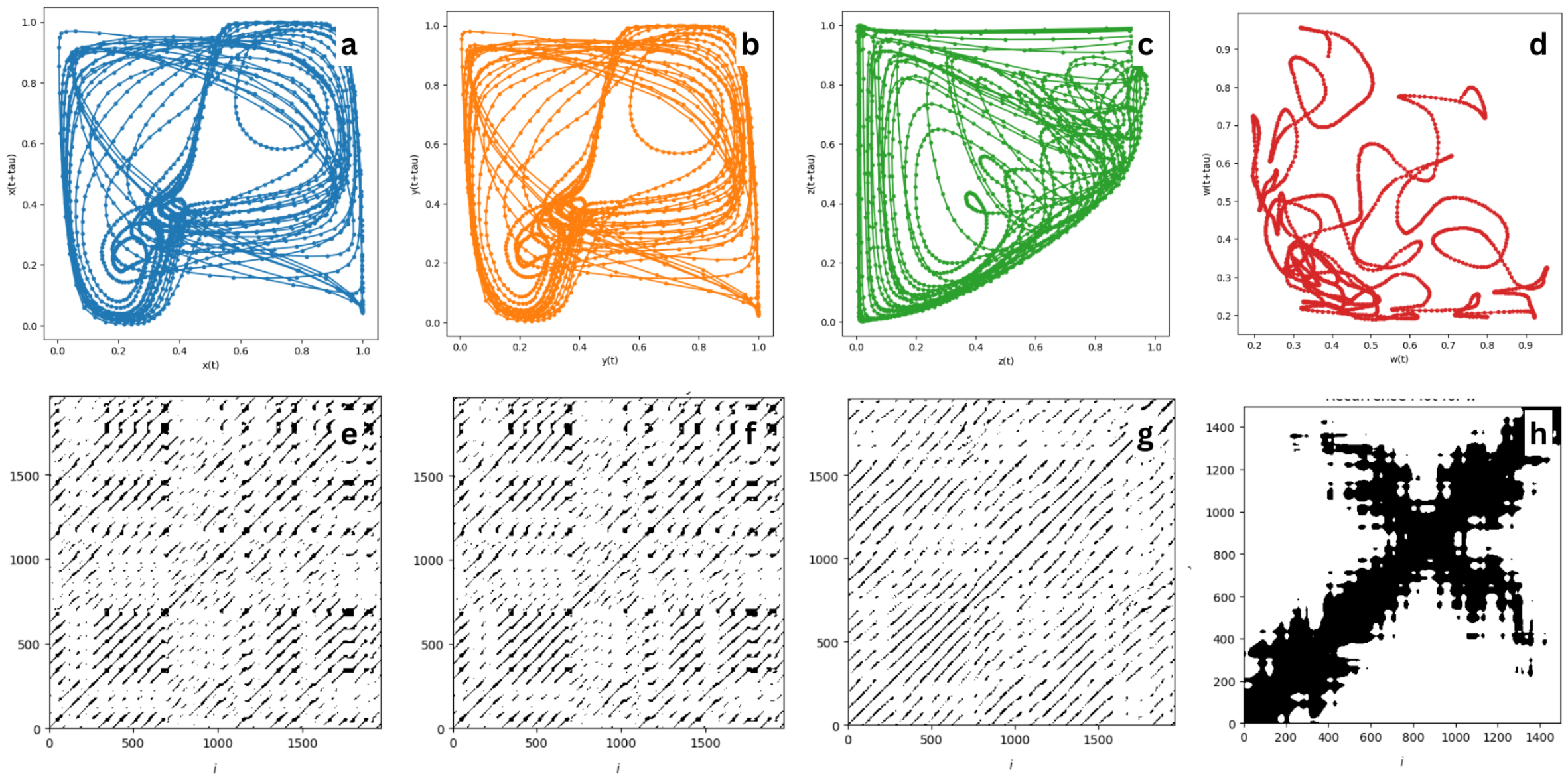}
    \caption{Reconstructed phase space structure and the corresponding Recurrence Plots from data of the Chen system in chaotic region for (a,e) x-data, (b,f) y-data, (c,g) z-data and (d,h) w-data.}
    \label{fig:chen chaotic}
\end{figure}

\begin{figure}[H]
    \centering
    \includegraphics[width=1\linewidth]{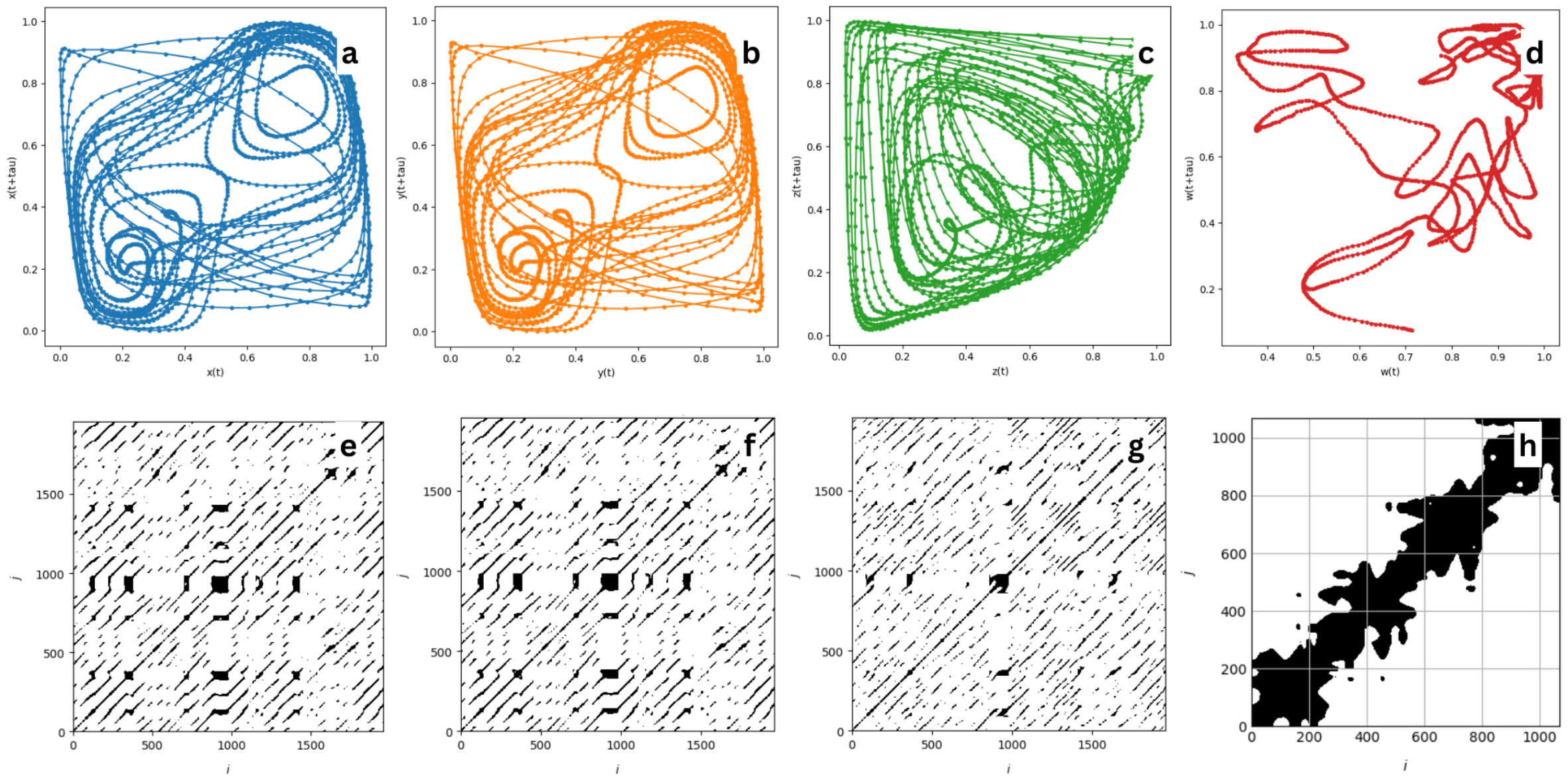}
    \caption{Reconstructed phase space structure and the corresponding Recurrence Plots from data of the Chen system corresponding to hyper chaotic state for (a,e) x-data, (b,f) y-data, (c,g) z-data and (d,h) w-data.}
    \label{fig:chen hyperchaotic}
\end{figure}

\begin{table}[h!] 
\scriptsize
  \centering
  
  \begin{tabular}{|>{\centering\arraybackslash}p{0.15\linewidth}|l|>{\raggedright\arraybackslash}p{0.3\linewidth}|>{\centering\arraybackslash}p{0.2\linewidth}|} \hline 
    
    Dynamical System& Dynamical Equations&  Parameter Values&Dynamical States of Behavior \\ \hline 
    
    Rössler System& 
    $\begin{aligned}
 & \dot{x}= - y - z \\
 & \dot{y}= x + ay \\
& \dot{z}= b + z( x - c))
\end{aligned}$ &  
    a = 0.2, b = 0.2, c = 5.7

a = 0.2, b = 0.2, c = 2.6 .
 & Chaotic 

Periodic \\ \hline 
    
    Lorenz System& $\begin{aligned}
& \dot{x} = p (y - x) \\
& \dot{y} = x (q - z) - y \\
& \dot{z} = xy - rz
\end{aligned}$
    &  p = 10.0, q = 28.0, r = 8/3 &Chaotic\\ \hline 
    
    Chen System \cite{gao_2009_analysis} & $\begin{aligned}
& \dot{x} = a (y - x) \\
& \dot{y} = -dx -xz + cy - w \\
& \dot{z} = xy - bz \\
& \dot{w} = x + k 
\end{aligned}$ &  
a = 36, b = 3, c = 22, d = 18, k = 0.

a = 36, b = 3, c = 20, d = 16, k = 0

a = 36, b = 3, c = 28, d = 24, k = 0.
 & Hyperchaotic

Periodic

Chaotic\\ \hline
    
  \end{tabular}
\caption{Details of the three dynamical systems and their parameter values used to generate synthetic data for periodic, chaotic and hyperchaotic dynamical states.}

\label{table:systems}
  
\end{table}
From the dynamical equations of each system, 20,000 data points are generated and then divided into 10 different segments of data sets.  From the trajectory points in the embedded phase space, the recurrence matrix and adjacency matrix are derived using Eqns.(\ref{eq:rec_mat}-\ref{eq:adj_mat}). 
We present the reconstructed phase space trajectories and their corresponding recurrence plots for typical segments of multivariate data from the Rössler, Lorenz, and Chen systems in Figs. (\ref{fig:rossler}-\ref{fig:chen hyperchaotic}). 


For the Rössler system, we find the data from the three variables reconstruct the same phase space structure. However, for Lorenz system, the embedded structure from x and y variables look similar but is very different from that of z variable.  As reported earlier \cite{10.1063/1.5049903,10.1063/1.1487570}, this can be seen as due to the inherent inversion symmetry in variables x and y, but not in z and so z-data cannot provide information about the symmetry of the system. Similarly the Chen system has symmetry with respect to inversion in x and y variables but not in z and w variables. This explains the difference in the reconstructed trajectories from z-data and w-data as compared to that from x-data and y-data.

The relevant measures are computed from their recurrence plots and recurrence networks for each segment of data and then averaged over  the 10 different data sets. The average values of the measures along with their standard deviations, \(\sigma\), thus obtained are presented in Tables.(\ref{tab:ross_lor_table}-\ref{tab:chen_table}) for data from each of the systems. The measures computed for white noise data of same size with \(\epsilon\) chosen as the \(8^{th}\) percentile of the distribution of its pairwise distances in the reconstructed 3-dimensional space are also shown . If the difference between the average measures from two datasets is more than three times the maximum \(\sigma\) among them, we consider their difference as significant or outside the confidence interval.

 We note that for data from Rössler system in the periodic sate the values of measures for the x-data, y-data, and z-data are within the significant levels. For data from its chaotic state, DET, and GCC values almost agree across the variables while ENT and CPL differ marginally for the y-data.

 \begin{table}[ht] 
\centering
\scriptsize

\begin{tabular}{| l  |l|l|l|l  |}
\hline
  &   \multicolumn{3}{|c|}{Rössler Periodic}& Noise\\
\hline
 &  x-data&y-data&z-data
& White Noise \\
\hline
DET &  0.988 ± 0.001&0.988 ± 0.001&0.990 ± 0.004& 0.235 ± 0.011\\
\hline
ENT &  
2.690 ± 0.123&2.839± 0.134&3.646 ± 0.120& 0.425 ± 0.042\\
\hline
CPL &  6.364 ± 0.060&6.369 ± 0.067&6.498 ± 0.061& 10.129 ± 0.374\\
\hline
GCC &  

0.757 ± 0.001&0.747 ± 0.001&0.750 ± 0.001& 0.355 ± 0.009\\
\hline

\end{tabular}
\caption{Recurrence measures obtained from data of x, y and z variables of  Rössler in the periodic state. The measures computed from white noise data are also shown for comparison.}
\label{tab:ross_p_wn}
\end{table}

We see that for the chaotic state of Lorenz system,  DET and CPL values for the x-data, y-data, and z-data are similar and fall within their respective confidence intervals. However, ENT and GCC values for the z-data differ from those of the x-data and y-data. 

\begin{table}[ht] 
\centering
\scriptsize

\begin{tabular}{| l  |l  |l  |l |l|l|l|}
\hline
  & \multicolumn{3}{|c|}{Rössler Chaotic
} & \multicolumn{3}{|c|}{Lorenz Chaotic
}\\
\hline
 & x-data& y-data& z-data& x-data& y-data&z-data
\\
\hline
DET & 0.883 ± 0.021& 0.934 ± 0.020& 0.900 ± 0.010& 0.915 ± 0.033& 0.871 ± 0.032&0.937 ± 0.015\\
\hline
ENT & 4.650 ± 0.175& 4.863 ± 0.236& 4.579 ± 0.104& 2.622 ± 0.121& 2.725 ± 0.025&3.364 ± 0.046\\
\hline
CPL & 3.798 ± 0.045& 3.680 ± 0.029& 3.774 ± 0.029& 4.590 ± 0.222& 4.486 ± 0.051&4.201 ± 0.091\\
\hline
GCC & 0.689 ± 0.012& 0.679 ± 0.011& 0.689 ± 0.019& 0.723 ± 0.011& 0.741 ± 0.001& 0.686 ± 0.001\\
\hline

\end{tabular}
\caption{Recurrence measures obtained from data of x, y and z variables of Rössler system and Lorenz system corresponding to the chaotic states. }
\label{tab:ross_lor_table}
\end{table}

For Chen system in the periodic state, the measures almost agree across the data from different variables. For data from the chaotic state,  DET for w-data is  lower than that of other variables while ENT and GCC are higher than that derived from data of the other variables. With data from the hyper chaotic state, we find ENT, CPL and GCC for w-data differ from  that of the other variables. 
 
\begin{table}[ht]
\centering
\scriptsize
\begin{tabular}{|>{\centering\arraybackslash}p{0.03\linewidth}|>{\centering\arraybackslash}p{0.05\linewidth}|>{\centering\arraybackslash}p{0.05\linewidth}|>{\centering\arraybackslash}p{0.05\linewidth}|>{\centering\arraybackslash}p{0.05\linewidth}|>{\centering\arraybackslash}p{0.05\linewidth}|>{\centering\arraybackslash}p{0.05\linewidth}|>{\centering\arraybackslash}p{0.05\linewidth}|>{\centering\arraybackslash}p{0.05\linewidth}|>{\centering\arraybackslash}p{0.05\linewidth}|>{\centering\arraybackslash}p{0.05\linewidth}|>{\centering\arraybackslash}p{0.05\linewidth}|>{\centering\arraybackslash}p{0.05\linewidth}|}\hline
 & \multicolumn{4}{|c|}{Hyperchaotic}& \multicolumn{4}{|c|}{Chaotic
} & \multicolumn{4}{|c|}{Periodic}\\\hline
\hline
 & x-data& y-data& z-data& w-data& x-data& y-data& z-data&w-data& x-data& y-data& z-data&w-data\\
\hline
DET & 0.911 ± 0.029& 0.885 ± 0.020& 0.864 ± 0.009& 0.651 ± 0.153& 
0.934 ± 0.004 & 0.927 ± 0.003 & 0.905 ± 0.008 &0.429 ± 0.152 
& 0.987 ± 0.003 & 0.986 ± 0.001 & 0.996 ± 0.001 &0.988 ± 0.003 
\\
\hline
ENT & 3.013 ± 0.095& 3.007 ± 0.046& 2.883 ± 0.058& 3.893 ± 0.300&   2.957 ± 0.033 & 2.899 ± 0.034 & 2.646 ± 0.047 &4.013 ± 0.419 
&  2.428 ± 0.260 & 2.503 ± 0.247 & 1.594 ± 0.057 &2.364 ± 0.203 
\\
\hline
CPL & 3.059 ± 0.061& 3.062 ± 0.058& 3.176 ± 0.076& 2.962 ± 0.450&   
3.102 ± 0.039 & 3.116 ± 0.038 & 3.182 ± 0.105 &2.091 ± 0.654 
&  6.650 ± 0.127 & 6.652 ± 0.127 & 6.448 ± 0.141 &6.646 ± 0.127 
\\
\hline
GCC & 0.645 ± 0.009& 0.651 ± 0.009& 0.648 ± 0.013& 0.716 ± 0.029&   0.642 ± 0.010 & 0.648 ± 0.010 & 0.652 ± 0.013 &0.830 ± 0.036 &  0.747 ± 0.001 & 0.747 ± 0.001 & 0.747 ± 0.001 &0.747 ± 0.001 \\
\hline

\end{tabular}
\caption{Recurrence measures obtained from data of x, y,  z  and w variables of  Chen system for the hyperchaotic, chaotic and periodic states.}
\label{tab:chen_table}
\end{table}
\subsection*{Detecting the dynamical states from multivariate data
}
The recurrent behaviour of a dynamical system as it evolves in  phase space is distinct between stochastic and non-linear deterministic  systems. Also the pattern of recurrences for the regular or periodic dynamics is different from that of chaotic or hyperchaotic behaviour.  By recreating dynamics from data or time series,  the recurrence pattern can be quantified using measures for RP and RN as discussed above. Hence these measures are conventionally used to discriminate between different dynamical states.  In this context in predicting the nature of dynamics, we may  benefit from using multiple data sets available from the same system.  We illustrate this by comparing the recurrence measures from data of different variables of the same system. The
four recurrence measures are compared among the data of different variables and across the three systems as well as with that from white noise in Figs (\ref{fig:det_ent}-\ref{fig:cpl_gcc}) below.
\begin{figure}[H]
    \centering
    \includegraphics[width=1\linewidth]{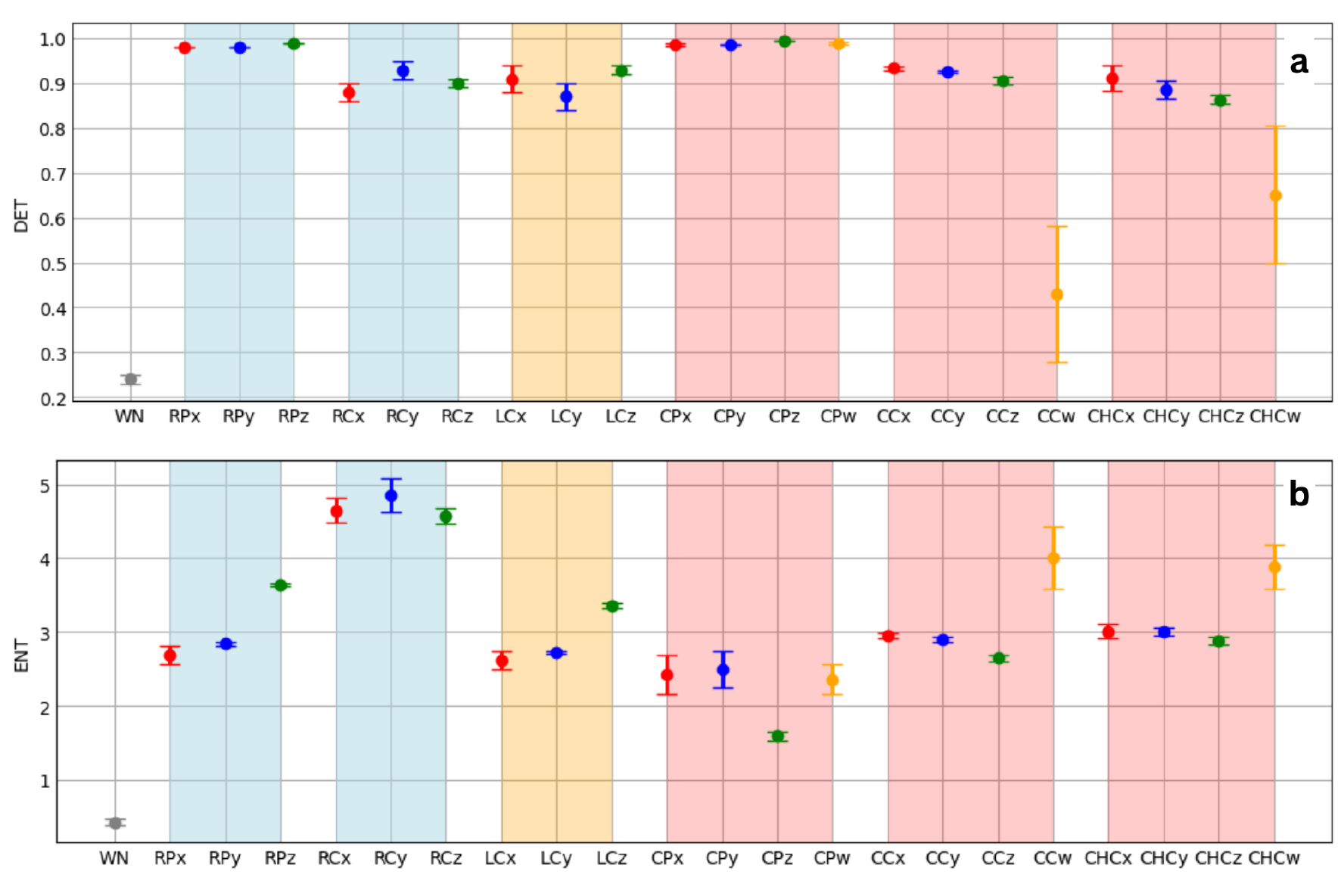}
    \caption{Recurrence plot measures Determinism and Entropy for data from white noise data (wn) and from the three dynamical systems, Rössler , Lorenz and Chen. Here RPx-z correspond to x-data, y-data and z-data of  Rössler system in the periodic sate, while  RCx-z, for data of Rössler system in chaotic state.  For Lorenz system in chaotic sate, LCx-z, stand for the data from x,y and z variables.  Similarly CPx-w correspond to data from periodic state of Chen system, CCx-w, data from its chaotic state and CHCx-w represent data from its hyperchaotic state.}
    \label{fig:det_ent}
\end{figure}

\begin{figure}[H]
    \centering
    \includegraphics[width=1\linewidth]{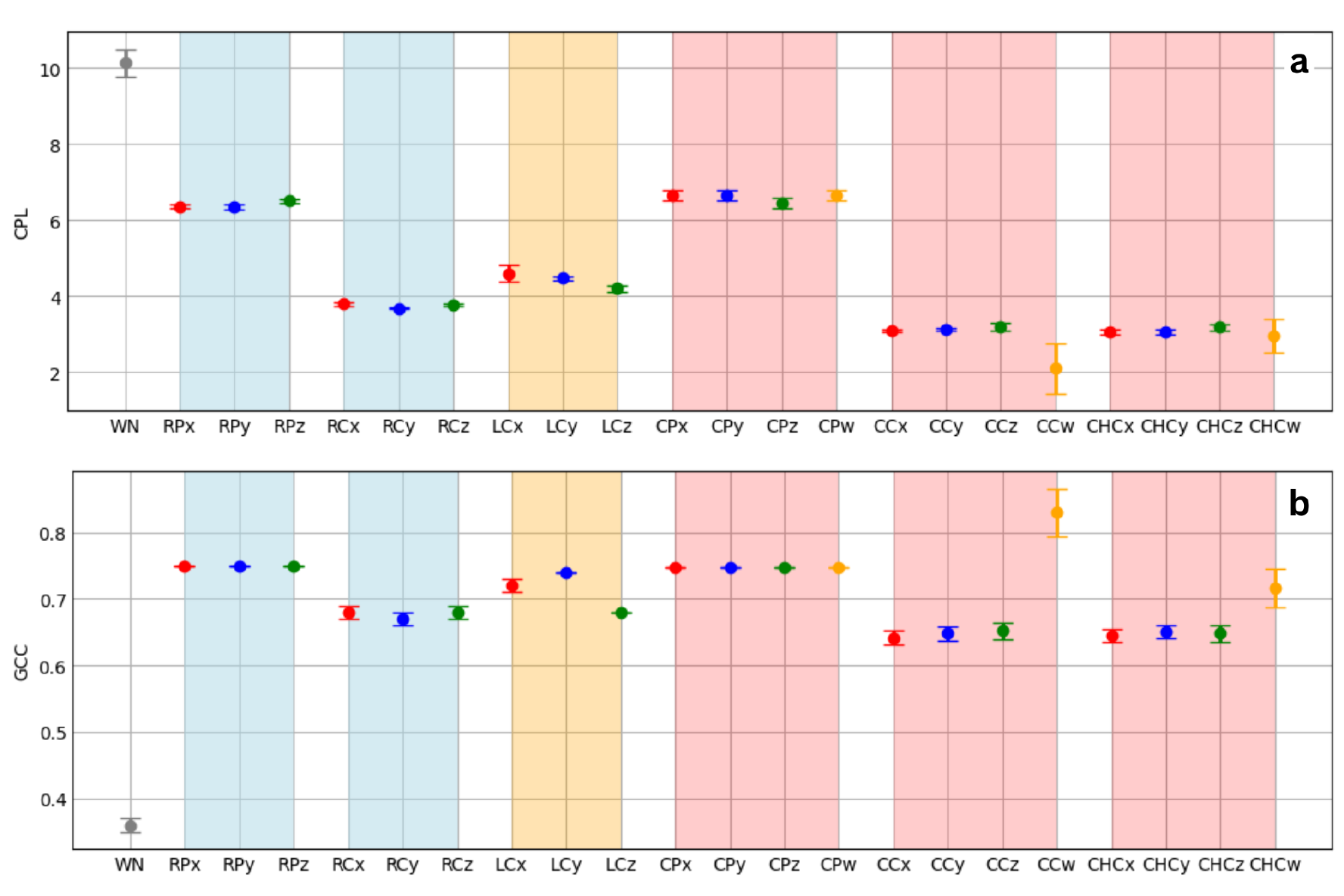}
    \caption{Recurrence Network measures, Characteristic Path Length and Global Clustering Coefficients  from white noise data (wn) and from the three dynamical systems, Rössler , Lorenz and Chen in different dynamical states. The symbols used on the x-axis are the same as in Fig 5.}
    \label{fig:cpl_gcc}

\end{figure}

 In general for the three standard systems studied, we find all the four measures from data of all the variables can be effectively used to distinguish deterministic and nonlinear nature from that of stochastic data.  However ENT and CPL are the best measures to differentiate between noisy data and chaotic or periodic data. 

For Rössler system, ENT from the y-data shows the maximum difference from that of the white noise for periodic state while data of all variables give similar differences in the chaotic case.  In the case of Lorenz system, for CPL the x-data gives the maximum difference from the white noise while ENT for the z-data shows the maximum difference. For Chen system in the chaotic state, ENT and GCC differ much from that of noisy data and the difference is maximum for w-data.  So also for the hyperchaotic state,  the difference in ENT and GCC is significant  for w-data. 

The same measures can  be used effectively to distinguish between the different dynamical states like periodic, chaotic and hyperchaotic.  In this application, for Rössler system, ENT again is a good measure as its value for chaotic data is higher when compared to that of periodic data,  while CPL for the periodic data is higher than that from chaotic data.  The difference in ENT is maximum for y-data but for CPL, the difference is more for z-variable. For data from Chen system, DET from w-data can be used to distinguish between chaos and periodicity, while we find ENT and CPL from w-data are also equally effective.  To detect chaos from hyperchaos, the GCC values from w-data can be most effectively used.

From above observations, we can say that  the recurrence measures computed from recurrence plot and recurrence networks vary across the different variables when their dynamical states are chaotic, or hyperchaotic and they differ to varying extents from the measures of noisy data.  

\section*{Conclusion}
We present a detailed study on the recurrence based measures derived from the reconstructed trajectories using the multivariate data generated from three standard systems, Rössler, Lorenz, and Chen. The practical efficacy of the approach in detecting the dynamics underlying the data is illustrated again in this study.  
While the four measures DET, ENT, CPL and GCC are useful in detecting deterministic dynamics from noisy data, we find ENT and CPL are the most effective to detect the nature of the dynamical state of the system. These measures are found to vary across the different variables when their dynamical states are chaotic, or hyperchaotic and they differ to varying extents from the measures of noisy data.  

It is possible that a particular dataset chosen for reconstruction does not contain enough information about the entire system to give reliable results. Thus, in many cases we may benefit from using data from all available observations of the system to improve the understanding of its dynamical evolution. The results of the computations in this study illustrate how different these measures are across data generated from the multiple variables of the same system. The main conclusions reached are valid, in general, regardless of the observable or the variable chosen for data. However, the choice of the variable can matter in extracting dynamical information from the embedded trajectory in some cases. The study thus reveals that the sensitivity of recurrence measures to system variables varies significantly across the studied systems and their dynamical states. This gives further support to the use of multiple embedding as an efficient way of extracting information from time series data when many simultaneous observations of the same system are available.

It is confirmed once again that the recurrence measures obtained from the reconstructed trajectory using data from any of the variables of a high dimensional system can detect its dynamical state and confirm if it is  different from noisy behaviour. To this extent, Taken’s embedding can be used with any of the multivariate data sets.  
We could isolate four measures that are robust and give consistent results even for short data sets. From all the different multivariate data sets analysed, we conclude ENT and  CPL are the most effective in detecting dynamical states from data. They also provide measures of the complexity of the dynamics and hence can be most effective in distinguishing periodic, chaotic and hyperchaotic states. The data from periodic state,  exhibit higher CPL and lower ENT values due to their simple, repetitive trajectories but for chaotic state,  they show lower CPL  and higher ENT values , reflecting their complex and less predictable behaviour. In the hyperchaotic state, the increased complexity and irregular nature lead to lower CPL and higher ENT values compared to chaotic or periodic states.

For Lorenz or Chen systems, the reconstructed trajectories from data of one or two of the variables look different from the original trajectories. We find the measures from their recurrences can still lead to similar conclusions in detecting the dynamics. However, the data from one or two of the variables may give measures that show more distinctions quantitatively and thus can be more effective in applications.  For this, we have to analyse data from all variables to identify the ones that are redundant and that which give maximum variations with noise and other possible dynamical states. This then gives an indication which measure to use to differentiate the nature of dynamics, and hence can be effective in applications using in machine learning  to detect the dynamics \cite{10.3389/fphys.2022.956320}. 

From our study we find that the recurrence measures are mostly similar for data across different variables for periodic state, while they differ for the chaotic and hyperchaotic states of the system. 
We further identify the ones that show significant differences among the variables. This must reflect the nonlinear relationship and symmetry among the variables that govern the dynamics of the system.  
Also this gives an indication of the particular variable whose data can be the best in detecting dynamics.

The study underscores the relevance of analyzing data from all  variables or observations of a complex system by illustrating how multiple datasets can be used to validate and improve predictions, and gain a more holistic understanding of the underlying dynamics. Thus the analysis of data from all the relevant variables together can complement and supplement the understanding about the dynamical processes that influence them and cause transitions by their changes.  This also highlights the implications of the study in understanding the relative effectiveness of multiple variables in revealing the underlying dynamical features. Thus when observational data from multiple channels or multiple wavelengths or energy bands are available for the same source, it is good to study the recurrence patterns from all of them so that we can understand the dynamical processes\cite{10.1063/1.4977950,MATHUNJWA2021102262} and their functional relationships that produce the data sets\cite{broadbent2023correlatedspectralrecurrencevariations,Ricketts_2023}. This can lead to future research directions in understanding the dynamics of complex systems from multivariate data sets.

\section*{Data Availability and Software Package}
All the data used in these studies are generated by integrating the ODE of the respective dynamical systems. The code used to analyse the data includes the publicly available package PyUnicorn; available at \url{http://www.pik-potsdam.de/∼donges/pyunicorn}, and frappy; available at \url{https://github.com/sgeorge91/frappy}) . Python scripts facilitated the combination of these packages in the analysis of this work.
\section*{Acknowledgments}
One of the authors SK acknowledges IISER Tirupati for computational facilities during the project work.

\nocite{*} 

\bibliographystyle{unsrt} 
\bibliography{sample} 

\end{document}